# Multiple Topological Magnetism in van der Waals Heterostructure of MnTe$_2$/ZrS$_2$

Zhonglin He, Kaiying Dou, Wenhui Du, Ying Dai[*], Baibiao Huang, Yandong Ma[*]

School of Physics, State Key Laboratory of Crystal Materials, Shandong University, Shandanan Street 27, Jinan 250100, China

*Corresponding authors: daiy60@sdu.edu.cn (Y.D.); yandong.ma@sdu.edu.cn (Y.M.)

**Abstract**

Topological magnetism in low-dimensional systems is of fundamental and practical importance in condensed-matter physics and material science. Here, using first-principles and Monte-Carlo simulations, we present that multiple topological magnetism (i.e., skyrmion and bimeron) can survive in van der Waals Heterostructure of MnTe$_2$/ZrS$_2$. Arising from interlayer coupling, MnTe$_2$/ZrS$_2$ can harbor a large Dzyaloshinskii-Moriya interaction. This, combined with ferromagnetic exchange interaction, yields an intriguing skyrmion phase consisting of sub-10 nm magnetic skyrmions under a tiny magnetic field of ~ 75 mT. Meanwhile, upon harnessing a small electric field, magnetic bimeron can be observed in MnTe$_2$/ZrS$_2$ as well, suggesting the existence of multiple topological magnetism. Through interlayer sliding, both topological spin textures can be switched on-off, suggesting their stacking-dependent character. In addition, the impacts of $d_\parallel$ and $K_{eff}$ on these spin textures are revealed, and a dimensionless parameter $\kappa$ is utilized to describe their joint effect. These explored phenomena and insights not only are useful for fundamental research in topological magnetism, but also enable novel applications in nanodevices.

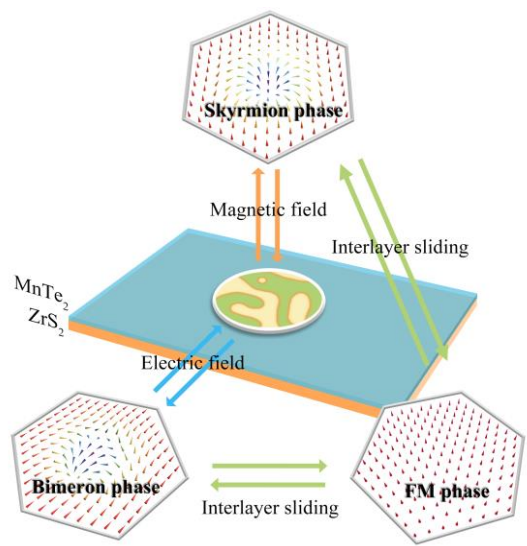

## Introduction

Since the first experimental observations of skyrmion lattice in bulk MnSi [1] and $Fe_{0.5}Co_{0.5}Si$ thin films [2], magnetic skyrmion, a typical topological magnetism, has been a prominent topic of condensed matter physics [3,4]. It is a spatially localized topological spin structure that is homotopically equivalent to a unit sphere, and characterized by quantized topological charge Q = ±1 [5]. Such non-trivial topological nature ensures an exceptional stability in terms of transition into trivial spin textures [e.g., ferromagnetic (FM) phase], making it technologically appealing for future memory and computing devices [6-8]. In addition to the study of magnetic skyrmions in perpendicularly magnetized systems, there has also been much effort in searching for new forms of topological magnetism in in-plane magnetized systems. One promising example is magnetic bimeron, which consists of a pair of merons and is of great interest recently because of its extraordinary properties [9,10]. In principle, magnetic bimeron is related to magnetic skyrmion with a π/2-rotation of each spin around in-plane axis [11]. As the topological charge Q is invariant under such rotation [5], magnetic bimeron exhibits topological nature and inherent stability as well [10].

The demand for device miniaturization in modern electronics stimulates the exploration of topological magnetism in two-dimensional (2D) materials [12,13]. The essential ingredient for the realization of topological magnetism is the Dzyaloshinskii-Moriya interaction (DMI) that exists under broken inversion symmetry and strong spin-orbit coupling (SOC). DMI is a form of antisymmetric exchange interaction, establishing a preferred chirality for spin textures [14-16]. In the past years, for realizing large DMI in 2D systems, extensive efforts have been devoted to Janus and ferroelectric structures [17,18]. While the former requires harsh conditions to realize in experiment [19,20], the latter severely limits the materials choices [21]. As a result, only a few 2D candidate systems have been proposed so far [16,18,22-25]. New and general mechanism for topological magnetism formation is highly desired for its exploration in the emerging area of 2D magnets. We also note that DMI can be induced in 2D van der Waals heterostructure (vdWH) [12], and superior to Janus and ferroelectric systems, vdWH systems exhibit higher experimental feasibility and tunability, providing an ideal platform for topological magnetism research. Nonetheless, the formation of topological magnetism in 2D vdWH is rarely reported [26,27], as the DMI is usually too weak to stabilize topological spin textures.

Here, through first-principles calculations and Monte-Carlo (MC) simulations, we propose the existence of multiple topological magnetism (i.e., skyrmion and bimeron) in 2D $MnTe_2/ZrS_2$ vdWH. Due to the strong interlayer interplay, $MnTe_2/ZrS_2$ possesses a large DMI. This, along with FM exchange interaction, results in the isolated zero-field magnetic skyrmion intrinsically in $MnTe_2/ZrS_2$.

When applying a tiny magnetic field of ~ 75 mT, intriguing skyrmion phase consisting of sub-10 nm magnetic skyrmions occurs. On the other hand, by harnessing a small electric field, magnetic bimeron can be observed. This confirms the existence of long-sought multiple topological magnetism in MnTe$_2$/ZrS$_2$. Furthermore, under interlayer sliding, both topological spin textures can be switched off, which suggests their stacking-dependent nature. Additionally, the roles of $d_\parallel$ and $K_{\text{eff}}$ in the formation of these spin textures are unveiled, and a dimensionless parameter $\kappa$ is utilized to characterize their joint effect. Our results greatly enrich the research of 2D topological magnetism.

**Results and Discussion**

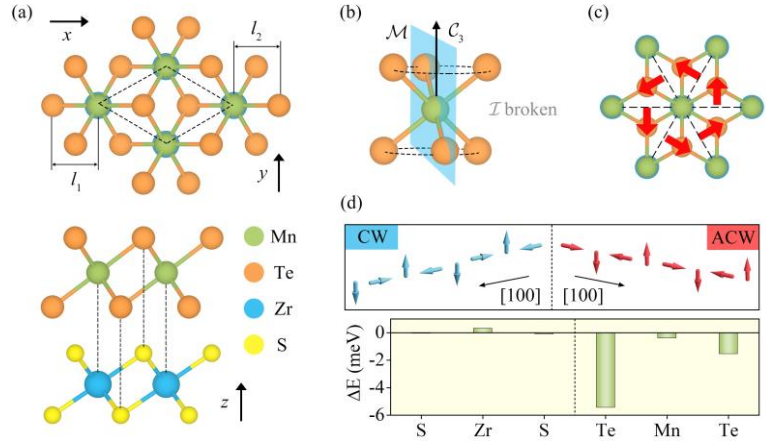

**Fig. 1**. (a) Top and side views of the crystal structure of MnTe$_2$/ZrS$_2$. The dashed diamond indicates the unit cell of MnTe$_2$/ZrS$_2$. (b) Distorted octahedral geometry of Mn atom. $\mathcal{M}$, $\mathcal{I}$ and $\mathcal{C}_3$ represent the ***M***-, ***I***- and ***C*$_3$**- symmetry, respectively. (c) DMI vectors ***D***$_{ij}$ (red arrows) between the nearest-neighboring Mn atoms. (d) Two spin-spiral configurations of CW and ACW employed to obtain DMI parameters and atomic-layer-resolved localization of the DMI associated SOC energy ΔE for MnTe$_2$/ZrS$_2$.

MnTe$_2$/ZrS$_2$ vdWH is composed of two layers of MnTe$_2$ and ZrS$_2$, which are known as FM metal and nonmagnetic semiconductor, respectively. These two constituent layers both belong to the 1T octahedral family of 2D transition metal dichalcogenides with space group P$\bar{3}$m1 (No. 164), thus exhibiting an inversion symmetry (***I***-symmetry). Obviously, when stacking these two layers together, ***I***-symmetry is broken. Here, we consider three typical stacking patterns of MnTe$_2$/ZrS$_2$ [see **Fig. S1**], among which the AA stacking pattern shown in **Fig. 1(a)** is proved to be the most stable configuration. In the following, we only discuss the AA stacking pattern of MnTe$_2$/ZrS$_2$. MnTe$_2$/ZrS$_2$ exhibits a distorted octahedral geometry ($l_1 \neq l_2$) for Mn atoms, resulting in the space group P3m1 (No. 156). To confirm the stability of MnTe$_2$/ZrS$_2$, we calculate its phonon spectra and perform ab initio molecular dynamics (AIMD) simulations. From **Fig. S2(a)**, we can see that all phonon branches are positive in the entire Brillouin zone, suggesting its dynamical stability. And as shown in **Fig. S2(b)**, after heating at 500 K for 5 ps, neither structure reconstruction nor bond breaking is found in MnTe$_2$/ZrS$_2$, confirming it is thermally stable.

Our calculations show that MnTe$_2$/ZrS$_2$ favors a spin-polarized state. The magnetic moment is calculated to be 3.35 $\mu_B$ per unit cell, which is mainly localized on Mn atom. For comparison, we also investigate the magnetic behavior of free-standing MnTe$_2$. The corresponding magnetic moment is calculated to be 3.23 $\mu_B$ per unit cell, which is in good agreement with the previous work [28]. To get further insight into the magnetic properties of MnTe$_2$/ZrS$_2$, we adopt the following atomically resolved Hamiltonian:

$$H = -J \sum_{<i,j>} (\boldsymbol{S}_i \cdot \boldsymbol{S}_j) - \lambda \sum_{<i,j>} (S_i^z \cdot S_j^z) - K_{\mathrm{CA}} \sum_i (S_i^z)^2 \\ - \frac{1}{2} \frac{\mu_0 g_0^2 \mu_B^2}{4} \sum_{\langle i,j \rangle} \frac{1}{r_{ij}^3} \left[ \boldsymbol{S}_i \cdot \boldsymbol{S}_j - \frac{3}{r_{ij}^3} (\boldsymbol{S}_i \cdot \boldsymbol{r}_{ij})(\boldsymbol{S}_j \cdot \boldsymbol{r}_{ij}) \right] \\ - \mu_{\mathrm{Mn}} B \sum_i S_i^z - \sum_{<i,j>} \boldsymbol{D}_{ij} \cdot (\boldsymbol{S}_i \times \boldsymbol{S}_j). \quad (1)$$

Here, $\boldsymbol{S}_i$ is unit vector ($|\boldsymbol{S}_i| = 1$) indicating the local spin of $i^{\mathrm{th}}$ Mn atom. $J$ represents the nearest-neighboring (NN) isotropic exchange interaction. $\lambda$ and $K_{\mathrm{CA}}$ describe the NN anisotropic exchange interaction and magnetocrystalline anisotropy (MCA), respectively. The forth term refers to the shape anisotropy $K_{\mathrm{SA}}$. This, combined with $K_{\mathrm{CA}}$, gives rise to an effective magnetic anisotropy $K_{\mathrm{eff}} = K_{\mathrm{CA}} + K_{\mathrm{SA}}$. $\boldsymbol{r}_{ij}$ is the vector from site $i$ to $j$, $\mu_0$ is the vacuum permeability, $g_0$ is the electron spin g-factor, and $\mu_B$ is the Bohr magneton. The penultimate term is Zeeman energy, where $\mu_{\mathrm{Mn}}$ and $B$ correspond to on-site magnetic moment of Mn atom and external magnetic field, respectively. $\boldsymbol{D}_{ij}$ characterizes the DMI vector for each pair of NN Mn atoms.

To obtain the magnetic parameters of $J$, $\lambda$ and $K_{\mathrm{CA}}$, we consider four different spin configurations of a 2×1 supercell of MnTe$_2$/ZrS$_2$ shown in **Fig. S3**. The isotropic exchange parameter $J$ is calculated to be 9.06 meV, which indicates a FM interaction between the NN Mn atoms. The underlying physics for such FM coupling in MnTe$_2$/ZrS$_2$ is associated with the Goodenough-Kanamori-Anderson rules [29-31]. In MnTe$_2$/ZrS$_2$, the Mn-Te-Mn bonding angle is 87°, approximately equal to 90°. According to the Goodenough-Kanamori-Anderson rules, the exchange coupling between NN Mn atoms is dominated by the superexchange interaction, leading to the FM coupling. This fact can also be confirmed by the spin charge density of MnTe$_2$/ZrS$_2$ shown in **Fig. S6(a)**. It can be seen that aside from the dominated distribution of spin-up charge densities on Mn atoms, there are some spin-down charge densities distributed on the adjacent Te atoms, which suggests the superexchange interaction between NN Mn atoms.

The NN anisotropic exchange parameter $\lambda$ is calculated to be 0.07 meV, which favors out-of-plane (OP) magnetization anisotropy. Besides $\lambda$, the easy magnetization axis (EMA) also relies on effective magnetic anisotropy $K_{\text{eff}}$, which, as mentioned above, contains two parts, i.e., magnetocrystalline anisotropy $K_{\text{CA}}$ and shape anisotropy $K_{\text{SA}}$. For 2D magnetic materials, $K_{\text{SA}}$ is only associated with the locations and magnetic moments of magnetic atoms, and usually prefers in-plane (IP) magnetization [32]. As expected, $K_{\text{SA}}$ is calculated to be -0.13 meV for MnTe$_2$/ZrS$_2$. For $K_{\text{CA}}$, it is calculated to be 0.36 meV, yielding the effective magnetic anisotropy $K_{\text{eff}}$ = 0.23 meV. Therefore, MnTe$_2$/ZrS$_2$ displays an OP EMA, and obviously, magnetocrystalline anisotropy plays a dominant role. To deeply understand this character, we investigate the underlying physics of OP magnetocrystalline anisotropy for MnTe$_2$/ZrS$_2$. Considering magnetocrystalline anisotropy is related to crystal field splitting and SOC, we introduce a perturbation theory of SOC effect, which can be written as [33]

$$\text{MCA} = \xi^2 \sum_{M,o,u,\sigma',\sigma} \sigma'\sigma \frac{\left|\langle u,\sigma|\hat{L}_z^M|o,\sigma'\rangle\right|^2 - \left|\langle u,\sigma|\hat{L}_x^M|o,\sigma'\rangle\right|^2}{E_{u,\sigma} - E_{o,\sigma'}}. \quad (2)$$

Here, $u$ and $o$ correspond to the unoccupied and occupied states of Mn atoms, respectively, $E_{u/o,\sigma}$ is the band energy of the state, and the spin indices $\sigma$ and $\sigma'$ run over ±1, referring to the two orthogonal spin states. M includes Mn and Te atoms. The positive (negative) value of MCA indicates OP (IP) magnetocrystalline anisotropy. **Fig. S4** presents the projected density of states for Mn-$d$ orbitals of MnTe$_2$/ZrS$_2$. Clearly, both $u$ and $o$ are mainly contributed by $|d_{xz}/d_{yz},\uparrow\rangle$ [↑ (↓) represents spin up (down) states], which enforces $\left|\langle u,\sigma|\hat{L}_z^M|o,\sigma'\rangle\right|^2 \gg \left|\langle u,\sigma|\hat{L}_x^M|o,\sigma'\rangle\right|^2$. As a result, positive value of magnetocrystalline anisotropy is obtained, giving rise to the OP magnetocrystalline anisotropy for MnTe$_2$/ZrS$_2$.

We then investigate DMI in MnTe$_2$/ZrS$_2$, which plays a vital role in establishing topological magnetism. The space group P3m1 of MnTe$_2$/ZrS$_2$ includes three mirror symmetries (***M***-symmetry) correlated by the three-fold rotation symmetry (***C*$_3$**-symmetry) [**Fig. 1(b)**]. According to Moriya's rule [34], the DMI vector can be expressed as $\boldsymbol{D}_{ij} = d_{ij,\parallel}(\boldsymbol{u}_{ij} \times \boldsymbol{z}) + d_{ij,z}\boldsymbol{z}$, where $\boldsymbol{u}_{ij}$ is the unit vector from site $i$ to $j$, and $\boldsymbol{z}$ is the OP unit vector. The two neighboring DMI of $\boldsymbol{D}_{ij}$ and $\boldsymbol{D}_{i(j+1)}$ are related through ***M***-symmetry, i.e., $\boldsymbol{D}_{i(j+1)} = \det(\boldsymbol{M})\boldsymbol{M}\boldsymbol{D}_{ij}$, which indicates that $d_{i(j+1),\parallel} = d_{ij,\parallel} =$ $d_\parallel$ and $d_{ij,z} = -d_{i(j+1),-z}$. This combined with $C_3$-symmetry results in a toroidal arrangement of the IP DMI component $d_{ij,\parallel}$ [see **Fig. 1(c)**] and a staggered arrangement of the OP DMI component $d_{ij,z}$ for the six NN Mn atoms. Concerning the toroidal arrangement of $d_{ij,\parallel}$, it potentially tends to produce the Néel-type magnetic skyrmion. While for the staggered arrangement of $d_{ij,z}$, it would lead to the vanishing of the OP DMI component in average, making little contribution for stabilizing magnetic skyrmion [17,25]. Therefore, only $d_\parallel$ is taken into consideration in the following. To obtain $d_\parallel$ of MnTe$_2$/ZrS$_2$, we consider the clockwise (CW) and anticlockwise (ACW) spin-spiral configurations, as shown in **Fig. 1(d)**. Intriguingly, it has a large intrinsic DMI of $d_\parallel$ = -1.39 meV.

For revealing the physical origin of such large DMI, the layer-resolved SOC energy difference ΔE between two spin-spiral configurations are calculated for isolated MnTe$_2$ and MnTe$_2$/ZrS$_2$. As shown in **Fig. S5** and **1(d)**, except for Te atomic layers, other atomic layers make minor contribution to DMI. This feature corresponds to the Fert-levy model [35]. For isolated MnTe$_2$, ΔE originated from two Te atomic layers are equal, but in opposite signs due to the protection of $M$-symmetry, leading to the vanishing DMI. In contrast, as shown in **Fig. S6(b)**, the strong electronic hybridization between the two layers in MnTe$_2$/ZrS$_2$ indicates a strong Te-S hopping, which results from large contact potential difference between MnTe$_2$ and ZrS$_2$ [**Fig. S6(c)**]. Apparently, the strong Te-S hopping has a significant impact on the lower Te atomic layer. That is to say, as illustrated in **Fig. 1(d)**, it inverses ΔE of the lower Te atomic layer, leading to the large DMI in MnTe$_2$/ZrS$_2$.

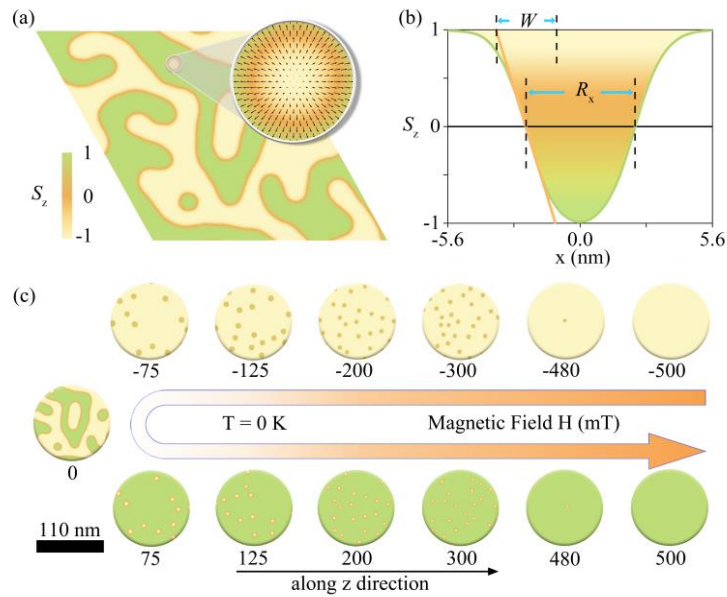

**Fig. 2**. (a) Spin texture of MnTe$_2$/ZrS$_2$ under zero-field. In (a), color map specifies the OP spin component and arrows indicates the IP spin component. (b) Line profiles of OP spin component along the central transect of magnetic skyrmion under magnetic field of ~ 75 mT. (c) Spin texture of MnTe$_2$/ZrS$_2$ as a function of external magnetic field.

Based on the magnetic parameters obtained above, the absolute value of the ratio between DMI vector and NN isotropic exchange interaction is calculated to be $\left|d_\parallel/J\right| = 0.15$. We note that $\left|d_\parallel/J\right|$ is usually considered as a criterion for the formation of magnetic skyrmion [17,25]: for $\left|d_\parallel/J\right| > 0.1$, magnetic skyrmion tends to be stable; while for $0 < \left|d_\parallel/J\right| < 0.1$, it usually indicates a trivial magnetic state. In this regard, magnetic skyrmion might exist in MnTe$_2$/ZrS$_2$. To verify this possibility, we investigate the topological spin textures in MnTe$_2$/ZrS$_2$ based on the parallel tempering MC simulations. Here, we employ the topological charge Q to characterize the magnetic skyrmion, which can be expressed as [36]:

$$Q = \frac{1}{4\pi} \sum_n q_n, (3)$$

with $\tan\frac{q_n}{2} = \frac{S_i^n \cdot (S_j^n \times S_k^n)}{1 + S_i^n \cdot S_j^n + S_j^n \cdot S_k^n + S_k^n \cdot S_i^n}$. Here, $S_i^n$, $S_j^n$, $S_k^n$ are the three spin vectors of the $n^{\text{th}}$ equilateral triangle in the anticlockwise lattice. **Fig. 2(a)** illustrates the spin textures of MnTe$_2$/ZrS$_2$. Obviously, a labyrinth domain pattern with Néel-type domain walls is observed. Along with these trivial patterns, intriguingly, a stabilized Néel-type magnetic skyrmion confirmed by Q = ±1 is also realized, without applying any external tuning.

We then investigate the evolution of spin texture of MnTe$_2$/ZrS$_2$ as a function of external magnetic field. As shown in **Fig. 2(c)**, the labyrinth domains shrink with increasing magnetic field, then disappear completely at a moderate magnetic field of ~ 75 mT. In this case, the Skyrmion phase is realized in MnTe$_2$/ZrS$_2$. More importantly, such phase is stable within a broad range of 75−480 mT. When further increasing the magnetic field, the Skyrmion phase is transformed into the trivial FM phase. **Fig. S7(a)** presents the radius of magnetic skyrmion as a function of magnetic field. It can be seen that the radius of magnetic skyrmion decreases with increasing the magnetic field from 0 to 75 mT. It is worthy emphasizing that the radius $R_x$ of magnetic skyrmion for MnTe$_2$/ZrS$_2$ is only ~ 4.5 nm under 75 mT, and the wall width $W$ is ~2 nm [see **Fig. 2(b)**]. With further increasing the magnetic field to 250 mT, the radius is almost unchanged. Upon increasing the magnetic field lager than 250 mT, the radius continues to decrease, and reduces to zero under 500 mT. In other words, all the Skyrmion phases in MnTe$_2$/ZrS$_2$ are composed of sub-10 nm magnetic skyrmions. Such a small size is highly promising for the practical applications in future skyrmionics devices. While for the density of magnetic skyrmions, as illustrated in **Fig. S7(b)**, a maximum of ~ $2\times10^{-3}$ per $nm^2$ (43 per supercell) is obtained under the magnetic field of 200-300 mT. By increasing the field further (300-480 mT), the density of magnetic skyrmion decreases rapidly, and suddenly shrinks to zero at 500 mT, corresponding to a quantum jump to the trivial FM phase.

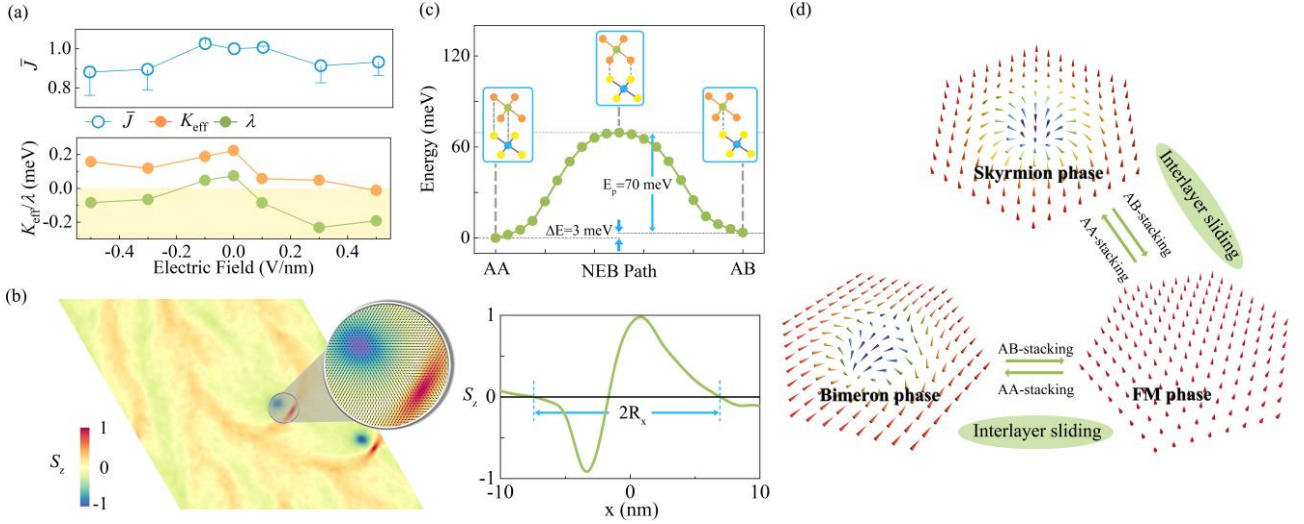

**Fig. 3**. (a) Magnetic parameters of $MnTe_2/ZrS_2$ as a function of applied electrical field. $\bar{J}$ is the ratio between $J$ under electric field and zero-field. (b) Spin texture of $MnTe_2/ZrS_2$ under electric field of 0.3 V/nm. Right panel in (b) illustrates the line profiles of OP spin component along the central transect of magnetic bimeron. In (b), color map specifies the OP spin component and arrows indicate the IP component. (c) Minimum energy path for interlayer sliding between AA and AB stacking patterns. (d) Schematic diagram of magnetic phase switching for $MnTe_2/ZrS_2$.

Similar to DMI, EMA, which is determined by $\lambda$ and $K_{eff}$, is also of great importance for establishing the topological magnetism. As unveiled above, $K_{eff}$ is sensitively depended on the electronic state around the Fermi level. In heterostructures, the electronic state can be easily manipulated via applying external electric field, which holds the possibility to tune $K_{eff}$ as well as the topological magnetism. To this end, we investigate the effect of electric field on the topological spin textures of $MnTe_2/ZrS_2$. **Fig. 3(a)** shows the magnetic parameters of $MnTe_2/ZrS_2$ as functions of external electric field. It can be seen that, with increasing the electric field from -0.5 to 0.5 V/nm, $J$ fluctuates in a narrow range, while $\lambda$ and $K_{eff}$ vary significantly. For the cases under electric field of -0.5 - 0.1 V/nm, the nature of OP EMA is preserved. Remarkably, when the electric field reaches 0.3 V/nm, the combined effect of $\lambda$ and $K_{CA}$ gives rise to an IP magnetization. The corresponding $d_\parallel$ under 0.3 and 0.5 V/nm is calculated to be ~ 1.29 and 1.50 meV, respectively. This results in $\left|d_\parallel/J\right|$

=1.55 and 1.76, respectively, which might stabilize IP topological magnetism.

We then perform the parallel MC simulations to explore the spin textures of MnTe$_2$/ZrS$_2$ under the electric field of 0.3 and 0.5 V/nm. As shown in **Fig. 3(b)**, under the electric field of 0.3 V/nm, the magnetic bimeron with Q = +1 is observed in MnTe$_2$/ZrS$_2$. Insect of **Fig. 3(b)** presents the OP magnetization of magnetic bimeron as a function of the radial coordinate. The radius of magnetic bimeron $R_x$ is ~ 14.3 nm, which is also suitable for practical application. We wish to point out that, in contrast to most of the previous work on bimeron [22,23,26], the magnetic bimeron achieved in MnTe$_2$/ZrS$_2$ is through electric field instead of magnetic field, indicating a more energy-saving way for exploring IP topological magnetism. When further increasing the electric field to 0.5 V/nm, as shown in **Fig. S8**, the magnetic bimeron still can be observed, but become vagueness, and the isolated merons emerge.

It is interesting to note that both DMI and EMA can be determined by interlayer hopping in MnTe$_2$/ZrS$_2$. And the interlayer hopping would be affected by the stacking pattern. Taking AB pattern as an example, we then discuss the effect of stacking pattern on the magnetic properties of MnTe$_2$/ZrS$_2$. The corresponding calculated magnetic parameters are summarized in **Table. S1**. And we obtain $\left|d_\parallel/J\right|$ = 0.069 < 0.1, which implies a trivial FM phase. To confirm this behavior, we conduct the parallel MC simulations. **Fig. S9** shows the spin texture of AB stacked configuration. As is expected, it exhibits a trivial FM phase. This suggests that, by transforming MnTe$_2$/ZrS$_2$ from AA to AB stacking pattern through interlayer sliding, both magnetic skyrmion and bimeron would be destroyed. Once such interlayer sliding is feasible, as shown in **Fig. 3(d)**, the on-off switching of the topological magnetism can be achieved in MnTe$_2$/ZrS$_2$. **Fig. 3(c)** illustrates the minimum energy path for interlayer sliding between the stacking order of AA and AB. The energy barrier is estimated to be ~ 70 meV per unit cell, suggesting the feasibility of the interlayer sliding.

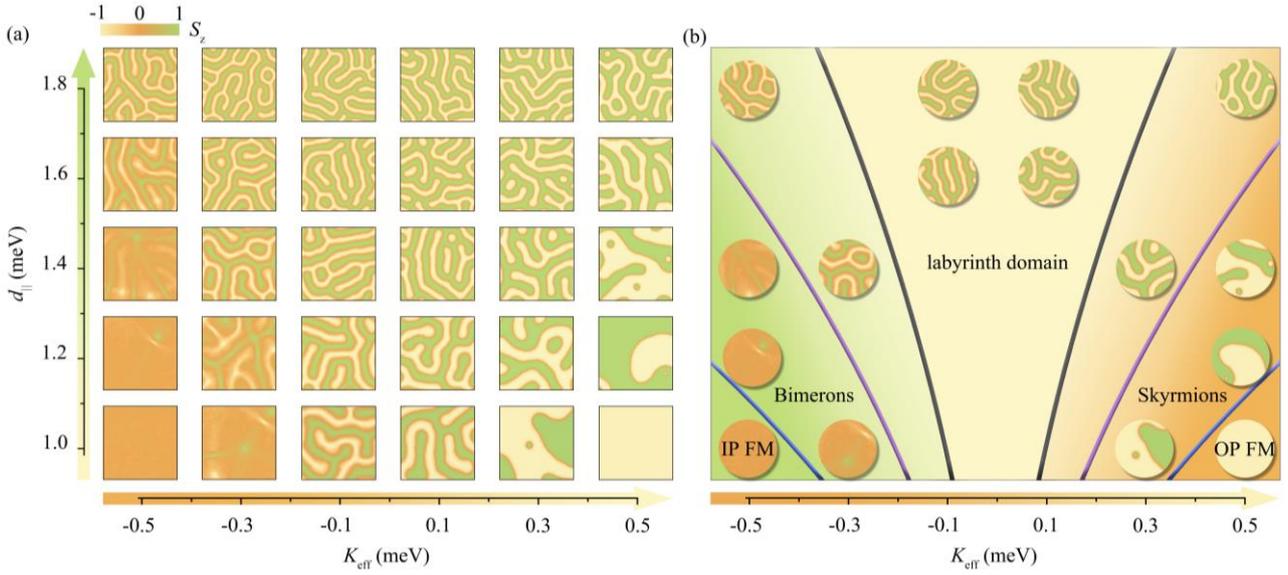

**Fig. 4**. (a) Spin texture diagram of MnTe$_2$/ZrS$_2$ as a function of $d_\parallel$ and $K_{eff}$. $J$ and $\lambda$ are set to 9.06 and 0 meV, respectively. (b) Spin texture diagram of MnTe$_2$/ZrS$_2$ described by $\kappa$. In (b), the gray, purple and blue lines represent $\kappa = \pm 1$, $\kappa = \pm 2$ and $\kappa = \pm 4$, respectively. Color map specifies the OP spin component

At last, a phase diagram of the spin texture as a function of $d_\parallel$ and $K_{eff}$ is summarized in **Fig. 4(a)**. Since $\lambda$ plays the same role as $K_{eff}$, we neglect the impact of $\lambda$, which is set to zero in MC simulations. As shown in **Fig. 4**, the trivial FM phase is observed under a small DMI and evolves into labyrinth domain phase with increasing DMI, which can shrink into Skyrmion phase upon introducing magnetic field. Therefore, a large DMI is essential for forming nontrivial topological spin textures. From **Fig. 4(a)**, with increasing DMI, we can also see that the density of labyrinth domain increases, and the size of labyrinth domain becomes smaller. Interestingly, arising from the large DMI, the magnetic skyrmions in MnTe$_2$/ZrS$_2$ would form a triangular lattice under magnetic field, leading to skyrmion crystal phase (**Fig. S10**). When modulating $K_{eff}$, as illustrated in **Fig. 4(a)**, the transition between the phases with magnetic skyrmion and bimeron can be observed. And the radius and density of magnetic skyrmion/bimeron are related to $|K_{eff}|$. Namely, a large $|K_{eff}|$ will introduce small density and radius of magnetic skyrmion/bimeron.

From above, we can see that both $d_\parallel$ and $K_{eff}$ have significant impacts on the spin textures of MnTe$_2$/ZrS$_2$. To describe their joint effect, we introduce a dimensionless parameter $\kappa$, which can be written as [37-39]:

$$\kappa = \left(\frac{4}{\pi}\right)^2 \frac{2JK_{eff}}{3d_\parallel^2}$$

The spin texture diagram of MnTe$_2$/ZrS$_2$ described by $\kappa$ is illustrated in **Fig. 4(b)**. Clearly, when $|\kappa| < 1$, labyrinth domain phase can be formed. For the spin textures around $|\kappa| = 1$, magnetic skyrmion appears close to the labyrinth domains. As $|\kappa|$ increases larger than 1, the size of labyrinth domain becomes larger, and the magnetic bimeron is observed. And with increasing $|\kappa|$, the magnetic skyrmion/bimeron moves further from the labyrinth domains, and the density and radius of magnetic skyrmion/bimeron decrease. When $|\kappa| \gg 1$, both magnetic skyrmion/bimeron and labyrinth domain decay into FM pattern, leading to the trivial FM phase; see **Fig. 4(b)**. It is important to stress that the dimensionless parameter $\kappa$ is only employed for describing physics of magnetic skyrmion in previous works [17,22], we show that it is also applicable for characterizing the cases of magnetic bimeron.

**Conclusion**

To summarize, we report the identification of multiple topological magnetism of magnetic skyrmion and bimeron in MnTe$_2$/ZrS$_2$ on the basis of first-principles calculations and Monte-Carlo simulations. Because of the strong interlayer hopping, this system presents a large DMI. The competition between the DMI and FM exchange interaction forms the spontaneous zero-field magnetic skyrmion in MnTe$_2$/ZrS$_2$. By introducing a tiny magnetic field of ~ 75 mT, skyrmion phase consisting of sub-10 nm magnetic skyrmions can be observed. Meanwhile, with including a small electric field, magnetic bimeron can be realized. This leads to the multiple topological magnetism in MnTe$_2$/ZrS$_2$. Moreover, through interlayer sliding, both topological spin textures can be switched off. Additionally, the roles of $d_\parallel$ and $K_{\text{eff}}$ on these spin textures are discussed, and a dimensionless parameter $\kappa$ is adopt to characterize their joint effect.

**Acknowledgement**

This work is supported by the National Natural Science Foundation of China (Nos. 12274261 and 12074217), Shandong Provincial Science Foundation for Excellent Young Scholars (No. ZR2020YQ04), Shandong Provincial Natural Science Foundation (Nos. ZR2019QA011), Shandong Provincial Key Research and Development Program (Major Scientific and Technological Innovation Project) (No. 2019JZZY010302), Shandong Provincial QingChuang Technology Support Plan (No. 2021KJ002), and Qilu Young Scholar Program of Shandong University.

**Data Availability**

The data that support the findings of this study are available within this article and its supplementary material.